\documentclass[final,5p,times,twocolumn]{elsarticle}

\usepackage{amssymb}
\usepackage{lipsum}
\usepackage{amsthm}
\usepackage{amsmath}
\usepackage{breqn}
\usepackage{cancel}
\usepackage{hyperref}
\usepackage{xcolor}

\biboptions{numbers,sort&compress,square,comma}

\journal{Physics Letters B}

\begin{document}

\begin{frontmatter}



\title{A perturbative geometric approach for photon spheres, massive particle surfaces and black hole shadows with mass variations}

            \author[first]{Oscar Lasso Andino\corref{oscar.lasso@udla.edu.ec}}
\affiliation[first]{organization={Escuela de Ciencias F\'{i}sicas y Matem\'{a}ticas, Universidad de Las Am\'{e}ricas},
            addressline={Redondel del ciclista, Antigua v\'{i}a a Nay\'{o}n}, 
            city={Quito},
            postcode={170504}, 
            state={},
            country={Ecuador}}
            \ead{oscar.lasso@udla.edu.ec}
            \cortext[oscar.lasso@udla.edu.ec]{Corresponding author}

\author[second]{Fernando G. Veloz}
\affiliation[second]{organization={Facultad de Ingeniería, Universidad San Sebastián },
            addressline={General Lagos 1163}, 
            city={Valdivia},
            postcode={5110693},
            country={Chile}}

\begin{abstract}
 The spacetime behavior at very extreme conditions, such as the regions near a black hole, can be very difficult to modelize. In this work we introduce a new geometric method that  allows to calculate the parameters of photon spheres, massive particle surfaces and shadow radius of black holes. We build upon a perturvative approach but bases in intrinsic curvatures, such as the geodesic and Gaussian curvatures. At leading order, the method allows to find  the radius of the perturbation in the time-like case, which has not been studied in the literature. In the null case we are able to recover the results found by the perturvative method only.  We also study the mass variations and how they influence the photon sphere radius and the massive particle surface radius, leading to a new and powerful result, that could provide new different research directions. The approach presented here will provide a set  of tools that will help to modelize gravity near extremely massive objects and help to improve the theoretical estimations of parameters than can  be tested in the next generation experiments.
\end{abstract}


\end{frontmatter}




\section{Introduction}
\label{introduction}
The recent experimental detection of black holes has significantly increased our understanding of gravity in extremal environments \cite{EventHorizonTelescope:2019dse,Luminet:1979nyg,Vagnozzi:2022moj,Falcke:1999pj}. The black hole shadow, formed when light goes around the horizon and hits the screen detector, can give a lot of information about the properties of the black hole, such as its mass. A plethora of black holes have been measured experimentally and interaction between them , such as black hole mergers have been detected \cite{Olejak:2019pln,Mroz:2024mse,LIGOScientific:2020kqk,LIGOScientific:2016dsl}, including the black hole which is at the center of our galaxy \cite{EventHorizonTelescope:2022wkp}. The advance in experimental techniques as well as the spectacular increase in new measurments is an indicator that  phenomema involving ultra massive objects needs to be modelized and new techniques that ease the calculations while giving new physical insights are needed \cite{Perlick:2021aok, Ayzenberg:2023hfw}. \\
What characterizes a black hole is the horizon, however, it cannot be measured directly, that is why the experiments focus on the near horizon effects, such as the shadow or the photon sphere. Recently, some works are pointing to a light ring detection \cite{Lupsasca:2024xhq, Gralla:2019xty, Wong:2024gph,Gralla:2020srx}, while this experimental detection is still under debate, the amount of analytical methods for studying these type of phenomena are increasing, giving a deep understanding of gravity under extremal conditions. In static spherically symmetric spacetimes, the location  of a light ring can be calculated by knowing the potential of the radial geodesic equation, then equating this potential and its derivative to zero, a radius of a light ring can be found. In \cite{Vertogradov:2024dpa} a new method for stuying light rings, at leading order, was presented.  The behaviour of the photon sphere and the shadow considering mass variations was studied. A spherically symmetric asymptotically flat black hole can be constructed by considering small perturbations of the Schwarzschild metric. Using this expansion the photon sphere and the shadow of different black holes can be calculated at leading order. This method has some advantages, the most important is to simplify the calculations that in principle can be really cumbersome. The time-like case has not been studied yet. This is where our work stands. We would like to follow a similar analysis but for time-like trajectories. However, the equations that we need to solve are complicated, therefore we follow a different direction.\\
A new geometrical method for studying circular geodesics was proposed in \cite{Qiao:2022hfv,Qiao:2022jlu}. In order to determine the location of a light ring, this  method uses the geodesic and the Gaussian curvature of the optical metric, which is a $2-$dimensional Riemannian metric built using the spacetime metric. Then, the condition for the existence of a light ring transforms to a condition over the geodesic curvature of circular orbits, namely the zeroes of this geodesic curvature represent the location of light rings. Similarly, stability is determined by the sign of the Gaussian curvature of the same optical metric. The geometric properties of the photon sphere that forms around a black hole can be studied  using the approach based on curvatures. One of the most evident advantage is the easiness that the method brings when studying time-like circular orbits \cite{Bermudez-Cardenas:2024bfi,Bermudez-Cardenas:2025hrp, Arganaraz:2021fwu, Cunha:2022nyw}. The purpose of this work is to join both methods, the perturbative  with the geometric one, for studying time-like trajectories and  the time-like counterpart of photon spheres, the so called massive particle surfaces. Using the geometric approach based on curvatures, when $m=0$, although  from a very different perpective, we recover the results obtained in \cite{Vertogradov:2024dpa}. However, our approach remains more general by including the time-like case. Moreover, it will allow us to study the stability of circular and null circular orbits. Contrary to what happens to the usual methods, we can also explore the geometric properties of massive particle surfaces and photon regions of spacetimes that are not asymptotically flat. Here, as an example, we estudy the Schwarzschild-AdS balck hole. Finally, an analysis of the radius of the photon sphere/massive particle surface, when the mass becomes dynamical, is performed. Our approach shows to be suitable for this kind of problems and also simplifies enormously the calculations, that even in the null case can be very long. \\
In section \ref{sec:2} we present a summary of conditions, based in curvatures, for the existence of light-rings and time-like circular orbits. In section \ref{sec:3} we study the null case. We recover the results in the literature but using the curvatures approach. We analyze the Reissner-Nordstr\"{o}m  and Bardeen black holes. In section \ref{sec:4} we present the time-like case and study in detail the Reissner-Nordstr\"{o}m  and Bardeen black holes.  In section \ref{sec:6} the influence of mass variations on the radii of photons spheres is analyzed. The asymptotically AdS case is presented in Appendix 1. Finally, the discussion section is presented in \ref{sec:7}. 

\section{Curvature and asymptotically flat spacetimes}\label{sec:2}

We are going to consider spherically symmetric asymptotically flat spacetimes of the form
\begin{equation} \label{metric:1}
    ds^2=-f(r)dt^2+\frac{dr^2}{f(r)}+r^2d\Omega^2_2\,,
\end{equation}
where $d\Omega^2_2=d\theta^2+\sin^2(\theta)d\phi^2$ and $f(r)$ is a function that depends only on the radial coordinate and satisfies the property $\lim_{r\rightarrow \infty}f(r)=1$. There are different ways to characterize circular orbits, one of them is a recently proposed geometric approach developed in \cite{Qiao:2022hfv,Qiao:2025ojr,Bermudez-Cardenas:2024bfi,Arganaraz:2021fwu}. In this proposal the geodesic and Gaussian curvatures of a Riemannian metric, coming from the spacetime metric, are used to characterize circular orbits and its stability. The criteria for the existence of null/time-like circular orbits is given by
\begin{equation}\label{condg:1}
    \kappa_g(r_{ph})=0\,,
\end{equation}
where $\kappa_g$ is the geodesic curvature defined in \eqref{geocurv:1}. If equation \eqref{condg:1} has a solution $r_{ph}$ then a null/time-like circular orbit is located at $r_{ph}$. On the other hand, the stability of the circular orbit at $r_{ph}$  can be determined by the sign of the Gaussian curvature \eqref{gausscurv:1}, thus \cite{Qiao:2022hfv,Qiao:2025ojr}:
\begin{eqnarray}\label{condg:2}
    K(r_{ph})>0 &\Rightarrow & \text{The light ring at $r_{ph}$ is stable}\,,\nonumber\\
     K(r_{ph})<0 &\Rightarrow & \text{The light ring at $r_{ph}$ is unstable}\,. 
\end{eqnarray}

In the null case, a Riemannian metric can be constructed easily and it is known as the optical metric \eqref{opticm2d:1}. Similarly, in the time-like case a Riemannian metric can be built by projecting over surfaces of constant energy\footnote{Mathematically, it represents a dimensional reduction over the directions of the Killing vector $\partial_t$ of the spacetime metric.}, the so called Jacobi metric \eqref{jacobim:1}. In both cases, the conditions \eqref{condg:1} and \eqref{condg:2}  can be used without any restriction. The approach has been discused widely in the literature and it certainly provides a simple geometrical way to characterize circular orbits. The advantages of the method will be made clear specially when time-like circular orbits are studied. Let us begin with the null case.

\section{The null case}\label{sec:3}
We are going to study the light-like perturbations using the geodesic and Gaussian curvatures. See appendix \ref{app:2} for detailed definitions. For null trajectories we have to use the optical metric \eqref{optic:1} coming from \eqref{metric:1}, then the geodesic curvature \eqref{geocurv:1} becomes 
\begin{equation}\label{geocurvnull}
  \kappa_g(r_{ph})=\frac{1}{\mathcal{E}}\left.\left(2f-rf'\right)\right|_{r=r_{ph}}\,.  
\end{equation}
Similarly, the Gaussian curvature \eqref{gausscurv:1} can be written using \eqref{optic:1} and \eqref{metric:1}, leading to \cite{Qiao:2022jlu,Qiao:2022hfv}
\begin{equation}
\mathcal{K}(r_{ph})=-\frac{f}{2\mathcal{E}^2r}\left.\left(r(f')^{2}-f(f'+rf^{''})\right)\right|_{r=r_{ph}}\label{gaussm:2}.
\end{equation}  
Due to the fact that both curvatures depend solely on $f(r)$, any perturbation in $f(r)$ will induce a perturbation in both curvatures. In the following sections we are going to show how these curvature perturbations can be calculated.

\subsection{Geodesic curvature perturbations and light rings}
We want to expand the function $f(r)$ around $r_{ph}$, where the photon sphere is located. In order to do that we consider the Schwarzschild metric together with an expansion that introduces small corrections to the Schwarzschild factor, thus \cite{Vertogradov:2024dpa}
\begin{equation}\label{expan:1}
f(r)=1-\frac{2M}{r}+\sum_{i=2}^{n}\frac{\beta_{i}}{r^{i}}+\mathcal{O}\left(\frac{1}{r^{n+1}}\right)\,, \hspace{0.8cm} \beta_{i}\ll 1\,.
\end{equation}
Note that $\lim_{r\rightarrow \infty}f(r)=1$, therefore the metric \eqref{metric:1} remains asymptotically flat. The radial is expanded as
\begin{equation}\label{rph}
r=r_{1}+\sum_{i=2}^{n} \beta_{i}r_{i}=r_{1}+\delta r\,,
\end{equation} 
where $r_{1}=r_{ph}$.\\

The Taylor expansion of geodesic curvature \eqref{geocurvnull} around the point $p=(r_{1}=3M,\beta_i=0)$ is written

\begin{equation}\label{geoexpan:1}
    \begin{split}
        k_{g}(r,\beta)&=\kappa_{g}(r_{1},0)+\left.\partial_r\kappa_{g}\right|_{p}(r-r_{1})+\sum_{i=2}^{n}\left.\partial_{\beta_i} \kappa_{g}\right|_{p}(\beta_{i}
        -0)\\
        &=\kappa_{go}+\partial_r \kappa_g|_{p}(r-r_{1})+\sum_{i=2}^{n}\beta_{i}\partial_{\beta_i} \kappa_g|_{p}\,.\\
    \end{split}
\end{equation}
 We want to calculate the radius of the perturbation $\delta r=r-r_1$ at first order, hence we equate expression \eqref{geoexpan:1} to zero and get
\begin{equation}
\sum_{i=2}^{n}\beta_{i}\left(r_{i}\partial_r\kappa_g+\partial_{\beta_i}\kappa_g\right)=0\,,
\end{equation}
where we have used \eqref{rph} and the fact that $k_{go}=0$. The linear independence of $\beta_i's$ leads to  
\begin{equation}\label{perturbr}
    r_{i}=-\left.\frac{\partial_{\beta_i}\kappa_g}{\partial_r \kappa_g}\right|_{p}\,.
\end{equation}
Equation \eqref{perturbr} is our first result. It gives the perturbation radii in terms of the derivatives of geodesic curvatures.  Using \eqref{expan:1} in \eqref{geocurvnull} we arrive to
\begin{equation}\label{geocurvexpan:1}
\kappa_g(r,\beta_i)=\frac{1}{\mathcal{E}}\left(2\left(1-\frac{3M}{r}\right)+\sum_{i=2}^{n}\frac{\beta_i(2+i)}{r^i}\right)\,,    
\end{equation}
hence, the derivatives of the geodesic curvature can be written:
\begin{eqnarray}
\frac{\partial \kappa_g}{\partial r}&=&\frac{1}{\mathcal{E}}\left(\frac{6M}{r^2}-\frac{\sum_{i=2}^{n}i(2+i)\beta_i}{r^{i+1}}\right)\,,\\
\frac{\partial \kappa_g}{\partial r}&=&-\frac{1}{\mathcal{E}}\sum_{i=2}^{n}\frac{2+i}{r^i}\,.
\end{eqnarray}
Finally, replacing the previous equations in \eqref{perturbr} we obtain
\begin{equation}\label{rad:1}
r_i=-\frac{(i+2)}{6M(3M)^{i-2}}.
\end{equation}
The corrections to the radii of Schwarzschild light rings $r_{ph}=3M$, inherited from the perturbations considered in \eqref{expan:1}, is given in \eqref{rad:1}, and it was first presented as the photon sphere theorem in \cite{Vertogradov:2024dpa}. We have derived, from geometric principles, the photon sphere theorem. Note that the result was expected because the condition for the existence of light rings used in \cite{Vertogradov:2024dpa} is exactly the condition \eqref{condg:1}. Similarly, by defining the impact factor $\sigma_o=L_o/E_o$ of the shadow of the black hole we can write $   \sigma_o^2=r^2_{ph}/f(r_{ph})$, which after using the condition for the existence of a light ring $\kappa_g=0$ it can be written  $   \sigma_o^2=2r_{ph}/f'(r_{ph})$. Thus, the shadow radius theorem can also be deduced.\\
The curvature approach can be used for studyng time-like circular orbits, where its power can be exploited to a great extent, generalizing the results already presented in \cite{Vertogradov:2024dpa}. Before going there, let us analize two examples: the Reissner-Nordstr\"{o}m and Bardeen black holes.

\subsection{Example: Reissner-Nordstr\"{o}m black hole}
The simplest perturbation is built by activating $\beta_2\neq 0$ with $\beta_{j}=0,\, j\geq 3$. If we set $\beta_2=Q^2$
 the expression \eqref{expan:1} becomes
 \begin{equation}\label{rnmetric}
     f(r)=1-\frac{2M}{r}+\frac{Q^2}{r^2}\,.
 \end{equation}
The previous function defines the metric of the Reissner-N\"{o}rdstrom black hole and has a photon sphere located in 
\begin{equation}\label{photrn}
    r_{ph}=\frac{3}{2}M+\frac{\sqrt{9M^2-8Q^2}}{2}\,.
\end{equation}
 
 \begin{figure}[h!]
    \centering
    \includegraphics[width=0.9\linewidth]{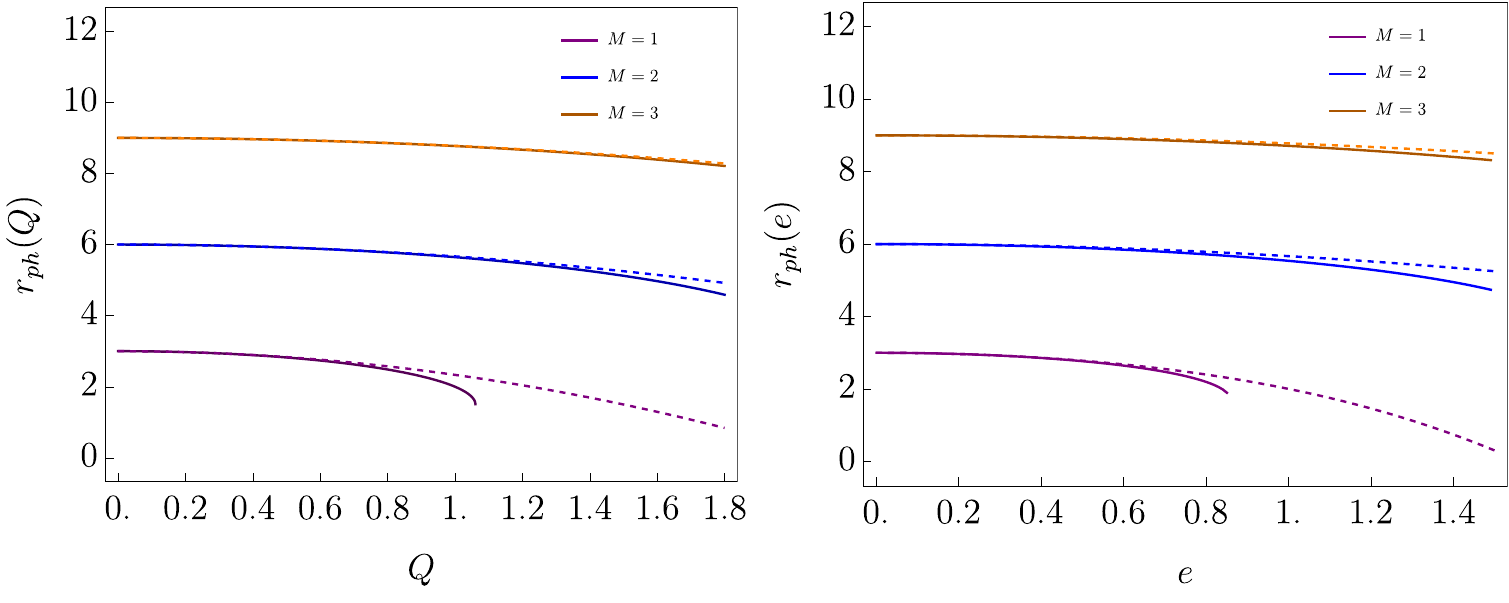}
    \caption{We have plotted  the photon sphere radius $r_{ph}$ as a function of $Q$ for  the Reisner-Nodstr\"{o}m black hole  (left panel), and  for the Bardeen black hole  with $e$(right panel). In both panels, the dashed curves correspond to the radius of the perturbed photon sphere, and the continuous line corresponds to the analytical result. We show three different cases: $M=1$ (purple curves), $M=2$ (blue curves) and $M=3$ (orange curves). }
    \label{fig:rnkg}
\end{figure}
In Fig.\ref{fig:rnkg} we have plotted the photon sphere radius of the Reissner-Nordstr\"{o}m metric \eqref{rnmetric} (left panel). The dashed curves  represent the radius calculated with the perturbation. For small values of $Q$ the approximation is good, but when the mass is bigger the zone where the approximations works well widens. Therefore, even for big values of $Q$ the approximation is good when big values of $M$ are also considered.

\subsection{Example: Bardeen  black hole}
We are going to analyze the Bardeen black hole \cite{Bardeen1968} whose line element is given by
\begin{equation}\label{bardeenf}
    f(r)=1-\frac{2Mr^2}{(e^2+r^2)^{3/2}}\,,
\end{equation}
where $M$ represents the mass of the black hole and $e$ is the monopole charge of a self-gravitating magnetic field sourced by non-linear electrodynamics. In order to find the radii of the perturbations we have to rewrite \eqref{bardeenf} as a series, hence the line function \eqref{bardeenf} can be approximated as
\begin{equation}\label{bardeenserie}
    f(r)\approx 1-\frac{2M}
{r}+\frac{3e^2M}{r^3}-\frac{15e^4 M}{4r^5}+\mathcal{O}(e^5)\,,
\end{equation}
therefore the coefficients of the perturbation \eqref{expan:1} are identified 
\begin{equation}
    \beta_3=3e^2 M,\hspace{1cm}\beta_5=-\frac{15e^4M}{4}\,.
\end{equation}
Equating the geodesic curvature \eqref{geocurvnull} to zero we are able find the radius of the photon sphere
\begin{equation}\label{rphb}
    r_{ph}=3M+\frac{35e^4}{216 M^2}-\frac{5e^2}{6}+\mathcal{O}(e^5)\,.
\end{equation}
In Fig. \eqref{fig:rnkg} we have plotted the radius of the photon sphere as a function of $e$ of  the  Bardeen black hole \eqref{bardeenserie} (right panel). The behaviour is very similar to Reissner-Nordstr\"{o}m, although the value of the charge $e$ where the approximation starts to fail is smaller than the equivalent value of $Q$.

\section{The time-like case}\label{sec:4}
In the time-like case, the Riemannian metric \eqref{opticm2d:1} used to calculate the curvatures is not useful anymore. Instead, we now use the Jacobi metric \eqref{jacobim:1}, which after replacing \eqref{metric:1}  becomes \cite{Bermudez-Cardenas:2024bfi,Bermudez-Cardenas:2025hrp}
\begin{equation}\label{condgeom:1}
\kappa_{g}(r_{tco})=\frac{1}{\mathcal{E}}\left(1-\frac{m^2}{\mathcal{E}^2}f\right)^{-3/2}\left.\left(2rf-r^2f'-\frac{2 m^2}{\mathcal{E}^2}r f^2\right)\right|_{r=r_{tco}}\,,
\end{equation}
where $m$ is the mass of the particle. If we set $m=0$ in \eqref{condgeom:1} we recover \eqref{geocurvnull}.  On the other side, the Gaussian curvature \eqref{gausscurv:1} becomes \cite{Bermudez-Cardenas:2024bfi,Bermudez-Cardenas:2025hrp,Cunha:2022nyw}
\begin{equation}\label{gaussianC}
\mathcal{K}(r_{tco})=-\left.\frac{f}{2\mathcal{E}^2 r\left(1-\frac{m^2}{\mathcal{E}^2}f\right)^2}\left(r (f')^2\left(\frac{1-\frac{2m^2}{E^2}f}{1-\frac{m^2}{\mathcal{E}^2}f}\right)-f (f'+r f'')\right)\right|_{r=r_{tco}}\,.
\end{equation}
When $m=0$ equation \eqref{gaussianC} reduces to equation \eqref{gaussm:2}. There is a relationship between both curvatures when evaluated at TCO's, the relationship reads \cite{Bermudez-Cardenas:2024bfi, Bermudez-Cardenas:2025duw}:
\begin{equation}\label{kmps}
    \mathcal{K}_{MPS}(r_{tco})=-\left.\frac{f^{1/2}(r)}{(\mathcal{E}^2-m^2f(r))^{1/2}}\partial_{r}\kappa_{g}(r)\right|_{r=r_{tco}}
\end{equation}
The previous expression will help us to study the time-like circular orbit perturbations and its stability only by knowing the behavior of the geodesic curvature $\kappa_g$ and its derivative.

\subsection{Geodesic curvature perturbations and TCO's}.
The expansion \eqref{expan:1} can be used to study perturbations of time-like circular orbits such as the innermost stable circular orbit (ISCO). We start by making a Taylor expansion of the geodesic curvature \eqref{condgeom:1} around $q=(r_{t_1}=6M,\beta_i=0,\gamma_1=m^2/\mathcal{E}^2_{1})$. Here, contratry to what happens in the null case, the quotient  $\gamma$ is also affected by the perturbation, hence 
\begin{equation}\label{gammapert}
\gamma=\gamma_{1}+\sum_{i=2}^n\beta_i \gamma_i\,,
\end{equation}
with
\begin{equation}
    \gamma_{1}=\frac{m^2}{\mathcal{E}_{1}^{2}}=\frac{r_{t_1}(r_{t_1}-3M)}{(r_{t_1}-2M)^2}=\frac{9}{8}\,,
\end{equation}
On the other hand, the geodesic curvature \eqref{geocurv:1} expansion becomes
\begin{equation}\label{geoexp:2}
    \begin{split}
       \kappa_{g}(r,\beta_i,\gamma)&=\kappa_{g}(q) +\left.\partial_r \kappa_{g}\right|_{q}(r-r_{t_1})+\left.\partial_{\gamma} \kappa_{g}\right|_{q}(\gamma-\gamma_{t_1})\\ 
&+\sum_{i=2}^{n}\left.\partial_{\beta_{i}}\kappa_{g}\right|_{q}(\beta_{i}-0)\\
        &=\kappa_{go}+\partial_r \kappa_g|_q\delta r+\partial_\gamma\kappa_g|_q\delta \gamma+\sum_{i=2}^{n}\beta_i\partial_{\beta_i}\kappa_g|_q\\
&=\left.\sum_{i=2}^{n}\beta_{i}\left(\gamma_i\partial_{\gamma}\kappa_g+\partial_{\beta_i}\kappa_g\right)\right|_{q}    
\end{split}\,,
\end{equation}
where we have used the fact that $\partial_{r}k_{g}(q)=0$ at the non-perturbed ISCO. 
The previous expression vanishes  when evalauted in the radius of a timelike circular orbit, therefore 
\begin{equation}\label{radius:2}
    \gamma_i=-\frac{\partial_{\beta_i}\kappa_g}{\partial_\gamma\kappa_g}\,.
\end{equation}
Now, we want an explicit expression depending on the parameter of the perturbation, then we  replace \eqref{expan:1} in \eqref{condgeom:1} and write
\begin{equation}
\begin{split}
    \kappa_{g}&=\frac{1}{2r\mathcal{E}}(1-\gamma f)^{-3/2}(2f-rf'-2\gamma f^2)\,,\\
    &=\mu\nu\sigma
    \end{split}
\end{equation}

where
\begin{equation}
    \mu=\frac{1}{2\mathcal{E}r}, \hspace{0.8cm} \nu=(1-\gamma f)^{-3/2},\hspace{0.8cm} \sigma=2f-rf'-2\gamma f^2.
\end{equation}
Using \eqref{expan:1} in the previous expressions we can write $\mu,\nu,\sigma$, evaluated at $q$, as a function of the expansion parameters:
\begin{equation}\label{abc:1}
    \mu\rvert_q=\frac{1}{12\mathcal{E}_{1}M},\hspace{0.6cm} \nu\rvert_q=6,\hspace{0.6cm}\hspace{0.6cm} \sigma\rvert_q=0.
\end{equation}
Using the results in \eqref{abc:1} we can calculate the coefficients of the series \eqref{geoexp:2}, thus
\begin{equation}\label{geoderivtl}
\kappa_{go}=0\,,\hspace{0.2cm}\partial_r\kappa_g=0\,,\hspace{0.2cm}
\partial_\gamma\kappa_g=-\frac{16}{27\mathcal{E}_{1}M}\,,\hspace{0.2cm}
\partial_{\beta_i}\kappa_g=\frac{4(i-1)}{\mathcal{E}_{1}(6M)^{i+1}}\,.
\end{equation}

Using the results in \eqref{geoderivtl}  we find that \eqref{radius:2} can be written as:
\begin{equation}\label{gammaper:1}
\gamma_{i}=\frac{9}{8}\frac{(i-1)}{(6M)^{i}}\,.
\end{equation}

The result  \eqref{gammaper:1}  shows that the perturbation \eqref{expan:1} induces a perturbation in the coefficients $\gamma_i$, which in its turn depend only on the mass of the Schwarzschild black hole $M$.  In the time-like case the quotient $\gamma=m^2/\mathcal{E}^2$ is also affected by the perturbation \eqref{gammapert}, therefore we have to calculate the corresponding factor. \\  
In order to workout the calculations we Taylor expand the derivatives of geodesic curvature around the point $q=(r_{t_1},0,\gamma_{1})$, thus

\begin{equation}\label{derivexpangeo:1}
\begin{split}
\partial_{r} \kappa_{g}&=\partial_r\kappa_{g}(q) +\left.\partial_r\left(\partial_r\kappa_{g}\right)\right|_{q}(r-r_{t_1})+\left. \partial _\gamma\left(\partial_r \kappa_{g}\right)\right|_{q}(\gamma-\gamma_{1})\\
&+\sum_{i=2}^{n}\left.\partial _{\beta_{i}}\left(\partial_r \kappa_g\right)\right|_{q}(\beta_{i}-0)\\
&=\partial_r\kappa_{g}(q)+\left.\partial_{rr}\kappa\right|_{q}\delta r+\left.\partial_{\gamma r}k\right|_{q}\delta \gamma+\sum_{i=2}^{n}\beta_{i}\left.\partial_{r\beta_{i}}\kappa_{g}\right|_{q}
\end{split}  
\end{equation}

We have to use that at the ISCO $\partial_r \kappa_g=0$, then expression \eqref{derivexpangeo:1} transforms, when equated to zero, to 
\begin{equation}
    \begin{split}
\partial _r \kappa_{g}=\sum_{i=2}^{n}\beta_{i}\left(r_{i}\partial_{rr}\kappa_{g}+\partial_{r\gamma}\kappa_{g}\gamma_{i}+\partial_{r\beta_{i}}\kappa_{g} \right)=0
    \end{split}
\end{equation}
hence
\begin{equation}\label{rtimelik}
 r_{i}=-\frac{\gamma_i\partial_{r\gamma}\kappa_g+ \partial_{r\beta_{i}}\kappa_{g}}{\partial_{rr}\kappa_g}
\end{equation}

Using \eqref{metric:1} and  \eqref{condgeom:1}, a quite long but straighforward calculation leads to
\begin{eqnarray}
    \partial_{r}\kappa_{g}(q)&=0\,,\hspace{1.5cm}  \partial_{rr}\kappa_{g}=&-\frac{1}{108\mathcal{E}_{1}M^3}\,,\\
     \partial_{\gamma r}\kappa_{g}&=-\frac{2}{9\mathcal{E}_{1}M^2}\,,\hspace{0.6cm}    \partial_{r\beta_{i}}\kappa_{g}=&\frac{-4i^2+9i-11}{\mathcal{E}_{1}(6M)^{i+2}}\,.
\end{eqnarray}
Therefore, the radii in \eqref{rtimelik} can be written
\begin{equation}\label{radtl}
 r_i=-\frac{(2i^2 +1)}{(6M)^{i-1}}\,.   
\end{equation}
Expression \eqref{radtl} is the radii of the perturbation, which is the time-like analogous to \eqref{rad:1}, and , as far we know, it has not been reported in the literature. 

\subsection{Example: Reissner-Nordstr\"{o}m black hole}
\begin{figure}[h!]
    \centering
    \includegraphics[width=1\linewidth]{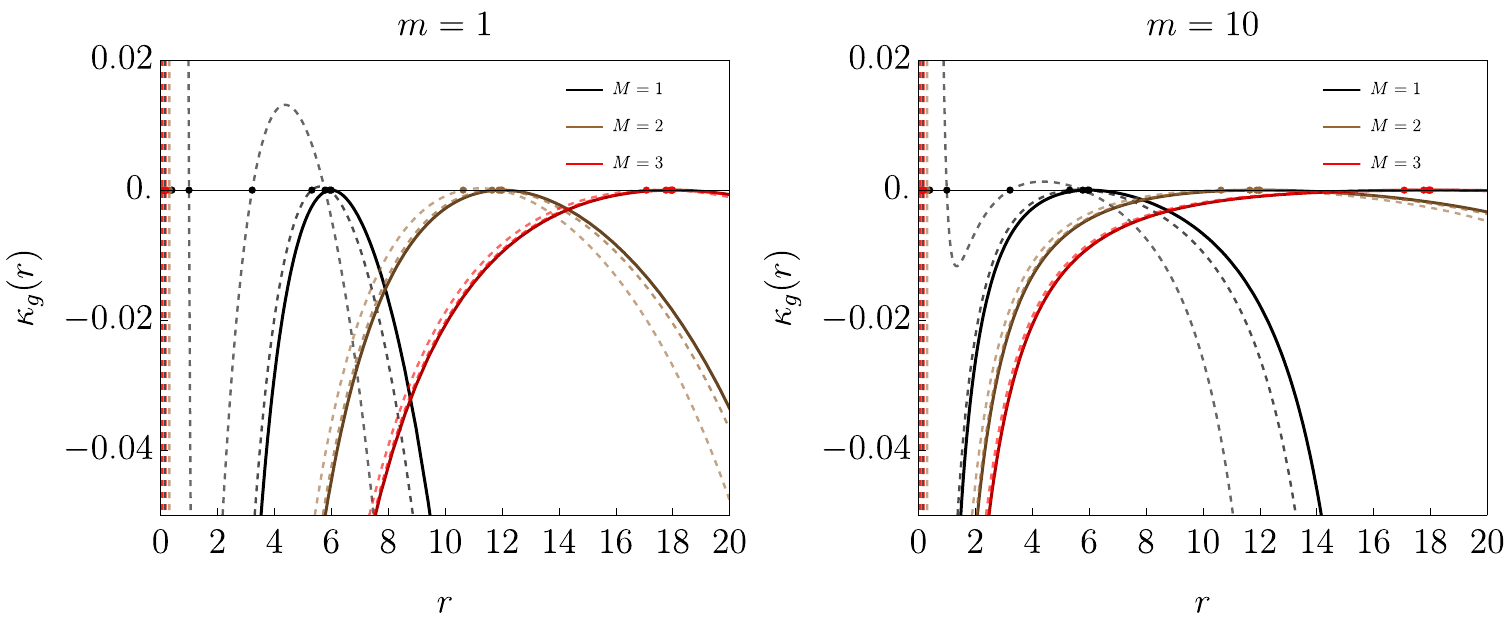}
    \caption{Geodesic curvature for time-like geodesics of the Reissner-Nordstr\"{o}m metric is plotted. In the left panel the geodesic curvature with $m=1$ is presented. The red curves correspond to $M=3$, the brown curves to $M=2$ and the black ones to $M=1$. The continuos curve for each case represent the geodesic curvature for time-like geodesics in the Schwarzschild case ($Q=0$). The dashed curves of each color are plotted with $Q=0.5,1$. In the right panel we have plotted the same curves but with $m=10$.}
    \label{fig:kgrntl}
\end{figure}

Let us show how the perturbative approach works for the Reissner-Nordstr\"{o}m  metric \eqref{rnmetric}. The geodesic curvature \eqref{condgeom:1} now depends on the mass of the particle that follows a time-like geodesic. The perturbations around $r_{t_1} =6M$ are given by \eqref{radtl}. The first conclusion is that the deviation from the radius of a time-like circular orbit does not depend on the charge $Q$, as in the null case, the radii depends only on the mass $M$ of the black hole. The geodesic curvature \eqref{condgeom:1} is calculated using \eqref{rnmetric} and is plotted in  Fig. \ref{fig:kgrntl}. In the left panel, geodesic curvature for the Schwarzschild metric ($Q=0$) for $m=1$ is represented by the continuous curves. The dashed curves represent the perturbations corresponding to Reissner-Nordstr\"{o}m metric ($Q=0.5,1$). We have considered the mass values $M=1,M=2,M=3$. In the right panel we have plotted the same curves but with $m=10$. The behaviour now is different compared with the null case, the geodesic curvature reachs a maximum which is locatted at zero, the point where $\partial_r\kappa_g=0$. Therefore, there is only one point where the geodesic curvature vanishes. This point corresponds to the location of the innermost stable circular orbit (ISCO). However, note the two zeros of the brown dashed curve, the approximation has failed. The charge is too big compared with $m$, that is why in the righ panel, when $m=10$, the maximum of the curve moves towards zero.

\begin{figure}[h!]
    \centering
    \includegraphics[width=0.8\linewidth]{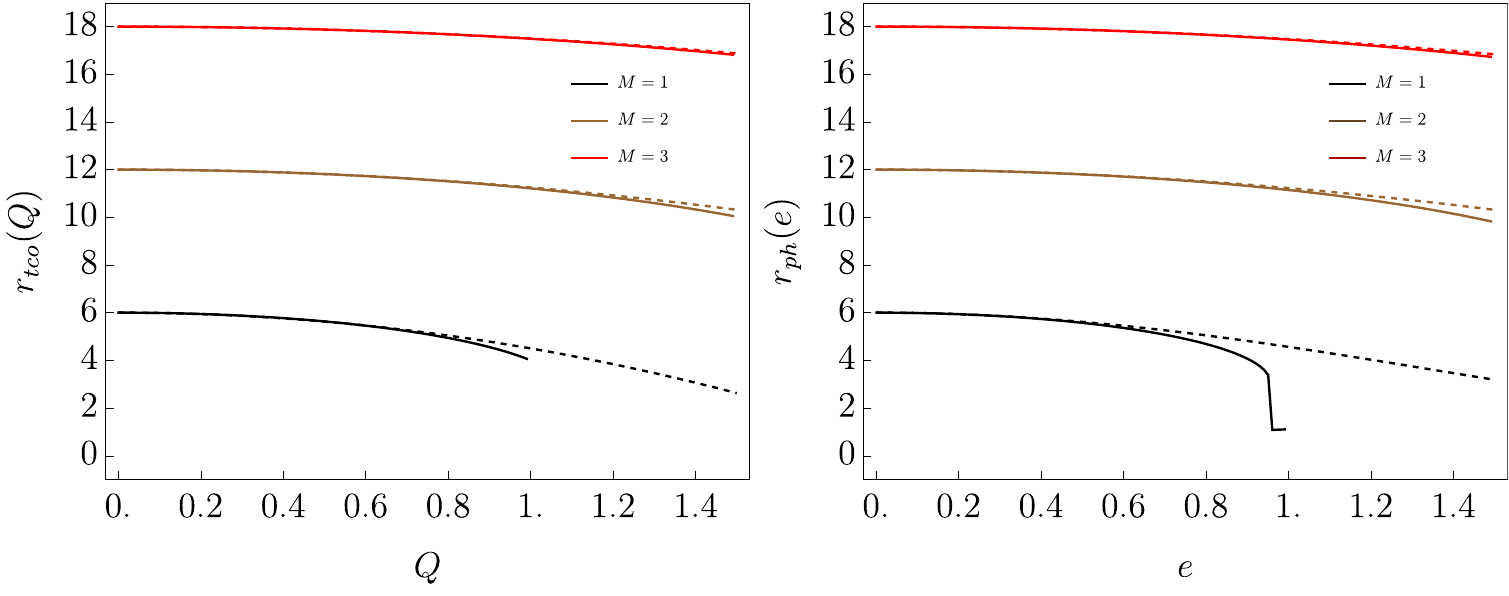}
    \caption{The plot shows the radius of the timelike circular orbit (continuous curve) and the approximatiation (dashed curve) as a function of charge $Q$ for the Reissner-Nordstr\''{o}m (left panel) and the Bardeen  (right panel) black holes. We have plotted the $M=1$ (black), $M=2$ (brown) and $M=3$ (red) cases.}
    \label{fig:rtcorn}
\end{figure}

In Fig. \ref{fig:rtcorn} the radius of a time-like circular orbits is plotted as a function of the charge $Q$. As it can be seen, the aproximmation is very good for small values of $Q$. However, this approximation will be improved when $M$ grows.

In Fig. \ref{fig:rtcorn} we have plotted the radius of a time-like circular orbit of the Bardeen black hole as a function of the charge $e$. The aprroximation (dashed curves) has a good agreement with the exact case when $e$ is small. However, as the mass $M$ of the black hole increases then the approximation improves,  in a similar way to the null case.

\subsection{Example: Bardeen black hole}
\begin{figure}[h!]
    \centering
    \includegraphics[width=1\linewidth]{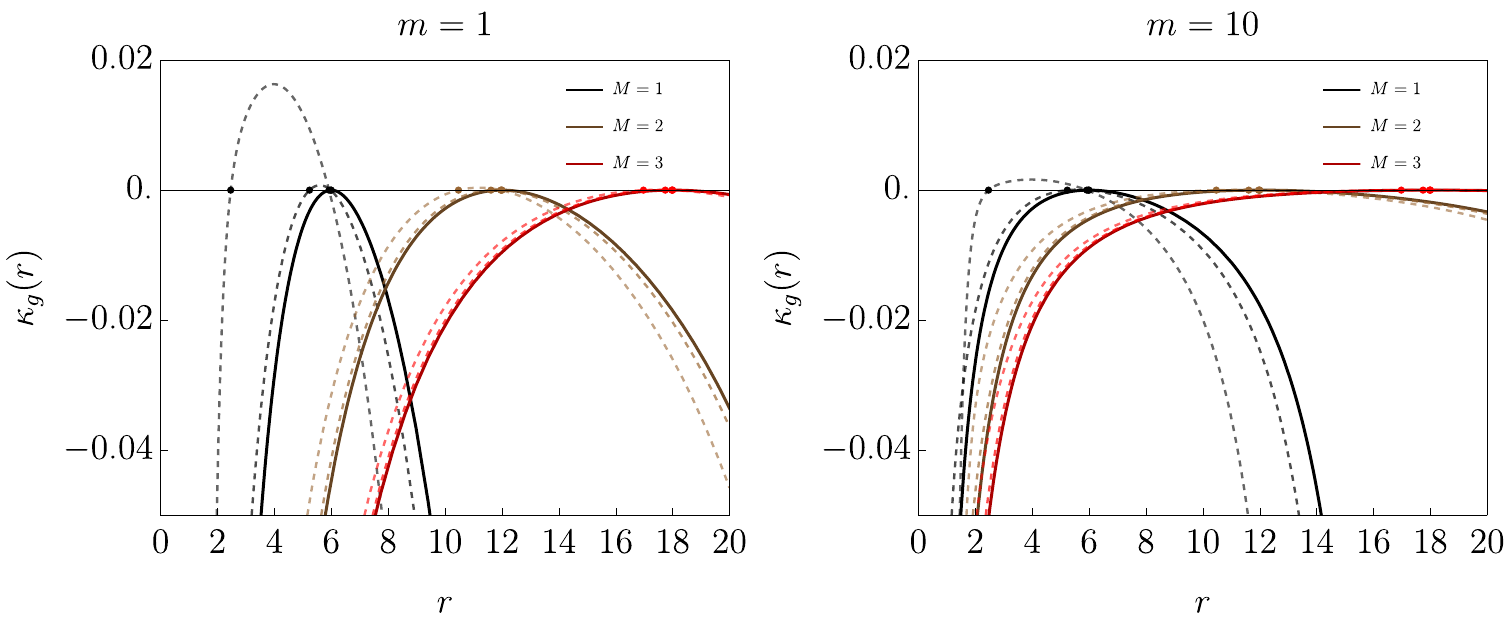}
    \caption{The geodesic curvatures for the Bardeen black hole are plotted. In the left plot we present the $m=1$ case and in the right plot  we present the $m=10$ case. Continuous curves represent the Schwarzchild case $e=0$, and the dashed curves the cases $e=0.5, 1$.}
    \label{fig:geocurvb}
\end{figure}

Here we are going to analyze the time-like circular orbits and their perturbations defined in the Bardeen black hole. As in the null case, we use the line element given by \eqref{bardeenserie} and calculate the geodesic curvature. In Fig. \ref{fig:geocurvb} we plotted the geodesic curvature for the bardeen black hole. In the left panel there is only one curve that has two zeros, all the other curves have a maximum which is located at $k_g=0$. Together with the vanishing of the geodesic curvature, we have that the derivative of that geodesic curvature also vanishes at the maximum points, which implies that the Gaussian curvature also vanishes, leading to the existence of a marginal stable circular orbits. In the left panel of Fig. \ref{fig:geocurvb} we show the $m=1$ case. It is direct to see how the geodesic reaches a global maximum at points where $\kappa_g$
vanishes. Similarly, in the right panel we have plotted $\kappa_g$ with $m=10$, then the mass of the test particle is bigger than $M$, then $\kappa_g$ has again a point where $\kappa_g$ and its derivative vanishes, a marginally stable circular orbit.

Finally, another advantage of the geometric approach based on curvatures is that it works quite well spacetimes are not asymptotically flat. See Appendix \ref{app:2} for the results on Schwarzschild-AdS.

\section{Photon sphere and massive particle surface radius with changing mass} \label{sec:6}
We are going to analyze a more general case, the one where the mass of the spacetime has a radial dependence. The line element is defined by
\begin{equation}\label{metric:r}
    ds^2=-f(r)dt^2+\frac{dr^2}{f(r)}+r^2d\Omega^2\,,\,\,\,\,\, f(r)=1-\frac{2M(r)}{r}\,,
\end{equation}
where $d\Omega^2=d\theta^2+\sin^2{\theta}d\phi^2$. We consider a small perturbation in the mass and the radius of the light ring, thus
\begin{equation}\label{massp}
    M(r)\rightarrow (1+\beta)M(r)\,,\,\,\,\,\, r_{phn}=r_{ph}+\beta r_{1}\,,\,\,\,\,\,|\beta|\ll 1\,,
\end{equation}
where $r_{ph}$ is the radius of the light ring before perturbation. For the null case we can calculate the radii of the perturbations  \eqref{perturbr}. Replacing \eqref{metric:r} in \eqref{perturbr} we get 
\begin{equation}\label{rmassv}
    r_1=\frac{r_{ph}}{1-2M'({ph})+M''(r_{ph})r_{ph}}\,,
\end{equation}
The previous equation was first obtained in \cite{Vertogradov:2024dpa}. We can identify the denominator of \eqref{rmassv} as the factor on the derivative of $\kappa_g$ \eqref{kmps} that determines the sign of the whole expression. Hence, we affirm that the sign of $r_1$ is given by $\mathcal{K}(r_{ph})$. In order to have a positive radius we need $\mathcal{K}(r_{ph})<0$ which according to the criteria \eqref{condg:2} correspond to a unstable null orbit \footnote{In order to show that the radius $r_1$ in \eqref{rmassv} is positive, the authors of \cite{Vertogradov:2024dpa} assume that the spacetime is supported by an specific isotropic energy momentum tensor, then by asumming a dominant energy condition and the existence of a bound from below for the radius of the photon sphere, then thy are able to show that the denominator of \eqref{rmassv}  is smaller than one, leading to the result that $r_1$ is always positive.}. The time-like case is more subtle. For unstable time-like orbits, one has  $\partial_{r}k_{g}> 0$, and the perturbed energy  \eqref{gammapert} associated with the new circular orbit becomes an additional quantity that must be taken into account. We can also consider the ISCO case. At this particular radius, we  have  to use \eqref{rtimelik} and write it as a function of $M(r)$.   Furthermore, at the ISCO we have that together with  $\partial_r\kappa_g=0$, equation \eqref{radius:2} has to be satisfied. However, the full expressions in the unstable orbits and the ISCO are not particularly illuminating as in the null correction \eqref{rmassv}. \\ 
In the null case, the mass and the radius of the photon sphere were realtes, if one increases the other do the same. In the time-like it is not the case. There could be very particular situations where the radius of the massive particle surface increases without increasing the mass of the black hole.
\section{Conclusions}\label{sec:7}
In this letter we have presented a new method for calculating the photon sphere radius, the  shadow radius and the massive particle surface radius of a static spherically symmetric black hole. First, we have derived the results presented in \cite{Vertogradov:2024dpa}   from a completely different geometric perspective. We have used a purely geometric method which takes into account the properties of intrinsic curvatures of a two dimensional Riemannian manifold. We recovered the expression for the radii of the perturbations, but now it is written using the geodesic curvatures, which represents our first important result. We have also given a simple calculation when the mass of the black hole changes dynamically, our method based in curvatures shows to be robust and allows interesting extensions. The dependence of the photon sphere radius on the variations of mass is a know result that we have derived from a different method. By  working out the case of Schwarzschild-AdS  black hole we have provided an extension to asymptotically AdS spacetimes.\\
The second part was devoted to the time-like case which, as far as we know, has not been studied in the literature, and where the advantages of the curvatures method are most clearly seen. We have shown how the radius of a massive particle surface, the time-like equivalent to the photon sphere, can be characterized perturbatively. As in the null case, we have provided an expression for the radii of the preturbations as a function of the geodesic curvature of the Jacobi metric. The time-like case is important when considering massive particles moving around the black holes, and it is related with the acreetion disk that can be formed around the black hole. The experimental test as well as the calculations, are far more involved when time-like trajectories are considered. The method presented here stands providing completely new results and open the way to future anlysis such as the dynamical formation of black holes. The results can be generalized to stationary-spherically symmetric black holes only by using a different Jacobi metric \cite{Arganaraz:2021fwu}, we lead this for future work. The analytical methods for analyzing photon spheres, massive particle surfaces and black hole shadows presented in this article can be used to study the existence of light rings or time-like circular orbits in different spacetimes, such as wormholes \cite{Arganaraz:2019fup,Duenas-Vidal:2022kcx,Bermudez-Cardenas:2025hrp}, providing a robust analytic way for attacking these kind of problems.\\
We hope that experimental arrays, such as the Event Horizon Telescope and all of its next generation devices will help to shed light on all these phenomena. The purely geometric approach presented here can be enhanced and tested. Hopefully, we can understand the phenomenology describing extreme gravity environments, such as black holes, providing a new perspective on theoretical and experimental aspects of black holes.

\section{Appendix 1. Curvature on symptotically AdS spacetimes}\label{app:1}
Another advantage of the geometric approach based on curvatures is that it works when spacetimes are not asymptotically flat. As an example, let us show how the calculations can be done when asymptotically AdS metrics are studied, see also \cite{LassoAndino:2026new}. Consider the following expansion 
\begin{equation}\label{expandads}
    f(r)=1-\frac{2M}{r}+\frac{r^2}{L^2}+\sum_{i=2}^{n}\frac{\beta_{i}}{r^i}\,, 
\end{equation}
When $\beta_i=0$ the radius of the light ring can be calculated using \eqref{condgeom:1}  and equating it to zero, leading to $r_1=3M$. Working the same procedure as in the asymptotically flat case it is direct to show that the raddii of the perturbation is given by \eqref{rad:1}, in other words, the result \eqref{rad:1} is not altered when the spacetime is asymptotically AdS. When $m\neq 0$ the procedure is the same but now the series is constructed around $r_{ta}$ which is a solution of the following equation  
\begin{equation}
     L^2M(r_{ta}-6M)+r_{ta}^3(4r_{ta}-15M)=0
\end{equation}
then using \eqref{expandads} and $u=(r_{1}=r_{ta}\,, \beta_i=0\,,\tilde{\gamma}_1)$ we get

\begin{equation}
    \kappa_{0}(u)=0\,,\hspace{0.5cm} \left.\kappa_{r}\right|_{u}=0\,,
\end{equation}
\begin{eqnarray}
    \left.\partial_{\gamma}\kappa_{g}\right|_{u}&=&-\frac{128(r_{ta}-3M)^{11/2}}{3\sqrt{3}\mathcal{E}_{1}M^{3/2}r_{r_{ta}}^{3}(4r_{ta}-15M)^2}\,,\\
    \left. \partial_{\beta_i}\kappa_{g}\right|_{u}
    &=&\frac{4\sqrt{r_{ta}-3M}}{3\sqrt{3}\mathcal{E}_{1}r_{ta}^{i+1}M^{3/2}}\left((i-2)(r_{ta}-3M)+3M\right)\,,
\end{eqnarray}
hence
\begin{eqnarray}
        \kappa_{g}(r,\beta, \gamma)&=&\sum_{i=2}^{n}\left \{ -\frac{128(r_{ta}-3M)^{11/2}}{3\sqrt{3}\mathcal{E}_{1}M^{3/2}r_{ta}^{3}(4r_{ta}-15M)^2} \gamma_{i}
        \right.\\
        & &
        \left. 
        + \frac{4\sqrt{r_{ta}-3M}}{3\sqrt{3}\mathcal{E}_{1}r_{ta}^{i+1}M^{3/2}}[(i-2)(r_{ta}-3M)+3M] \right \}\,.\nonumber
\end{eqnarray}
then, equating the previous equation with zero we get

\begin{equation}
\gamma_{i}=\frac{1}{32 r^{i-2}}\frac{(4r_{ta}-15M)^2}{(r_{ta}-3M)^5}\left((i-2)(r_{ta}-3M)+3M\right)\,.   
\end{equation}
In order to find the radii of perturbations in the time-like cases we have to use $\partial_{r} k_{g}=0$. Taylor expanding this relation and using \eqref{expandads} we get

\begin{equation}
\partial_{rr}\kappa_{g}=\frac{4(2r_{ta}-15M)\sqrt{r_{ta}-3M}}{\sqrt{3}\mathcal{E}_{1}\sqrt{M}r_{ta}^4} 
\end{equation}

\begin{equation}
    \partial_{\gamma r}\kappa_g=\frac{32(21M-8r_{ta})(r_{ta}-3M)^{9/2}}{3\sqrt{3}\mathcal{E}_{1}M^{3/2}r_{ta}^{4}(4r_{ta}-15M)^2}
\end{equation}

\begin{equation}
    \partial_{\beta_{i}r}\kappa_{g}=\frac{\mathcal{A}}{3\sqrt{3}\mathcal{E}_{1}M^{3/2}r_{ta}^{i+2}\sqrt{r_{ta}-3M}}\,,
\end{equation}
where
\begin{eqnarray}
    \mathcal{A}=-4(r_{ta}-3M)^2 i^2 +(16r_{ta}^2-117Mr_{ta}+207M^2)i\,,\nonumber\\
    -16r_{ta}^2 +114Mr_{ta}-207M^2\,.
\end{eqnarray}
Finally, the expression for the radii of the perturbation becomes
\begin{equation}\label{radads}
    r_{i}=\frac{\mathcal{B}}{6M(r_{ta}-3M)(2r_{ta}-15M)}\,,
\end{equation}
where
\begin{equation}
   \mathcal{B}=r_{ta}^{2-i}\left(2(r_{ta}-3M)^2 i^2 -4(r_{ta}-3M)(r_{ta}-6M)i+9M^2\right)\,.
\end{equation}
Equation \eqref{radads} correspsonds to the radii of the perturbation, in the time-like case, for Schwarzschild-AdS metric.

\section{Appendix 2. Curvatures, photon spheres and massive particle surfaces}\label{app:2}
Let us consider a $2$-dimensional surface $\mathcal{S}$ on which we have defined a coordinate system with coordinates $(X_1,X_2)$, then a metric defined over $\mathcal{S}$ is given by
\begin{equation}\label{met:1}
    ds^2=UdX_1^2+VdX_2^2\,.
\end{equation}
Any curve $\alpha(s)=(X_1(s),X_2(s))$ defined over $\mathcal{S}$ has a geodesic curvature that constitutes a measure of how far that curve is of being a geodesic. In particular, for a circular trajectory the geodesic curvature can be calculated as \cite{Berg:1988,Kre:1991}
\begin{equation}\label{geocurv:1}
    \kappa_g=\left.\frac{1}{2\sqrt{U}}\frac{\partial \ln (V)}{\partial r}\right|_{r=r_o}\,.
\end{equation}
where $r_o$ is the radius of the circular orbit. Hence, one way of characterizing a circular geodesic \footnote{The null trajectories of this type are known as light rings (LR's) and the time-like trajectories are known as time-lice circular orbits (TCO).} is the condition:
\begin{equation}\label{condgeo:1}
\kappa_g=0\,.
\end{equation}
If we want to measure how far a surface of being flat we can use the Gaussian curvature $\mathcal{K}$. The Gaussian curvature of the surface $\mathcal{S}$  can be calculated using the metric \eqref{met:1}, and it is given by \cite{Berg:1988,Kre:1991}: 
\begin{equation}\label{gausscurv:1}
    \mathcal{K}=-\frac{1}{2\sqrt{U V}}\left[\frac{\partial}{\partial x_1}\left(\frac{\partial_{x_1}V }{\sqrt{UV}}\right)+\frac{\partial}{\partial x_2}\left(\frac{\partial_{x_2}U }{\sqrt{UV}}\right)\right]\,.
\end{equation}
The Gaussian curvature $\mathcal{K}$ can be used to study the stability of the circular geodesics defined over $\mathcal{S}$. Only with the sign $\mathcal{K}$  we can analyze the stability of a orbit. $\mathcal{K}<0$ implies that the orbit is unestable and  $\mathcal{K}>0$ implies that the orbit is stable \cite{Qiao:2022jlu,Qiao:2022hfv}. Both instrinsic curvatures are calculated using Riemannian metrics.\\
Geodesic motion in spacetimes, which are Lorentzian, can be characterized using the geodesic and Gaussian curvatures of a two dimensional Riemannian metric \cite{Gibbons:2015qja, Das:2016opi, Arganaraz:2019fup,Duenas-Vidal:2022kcx}. 
Therefore, in order to study geodesic motion in Lorentzian spacetimes, the problem of finding a Riemannian metric associated with a spacetime that inherits its geodesic structure needs to be addressed. In the following section we present two methods known in the literature for calculating a Riemannian metric from a Lorentzian metric.
\subsection{The optic metric and the photon sphere}
The optical metric can be used to study the motion along null geodesics. Given a Lorentzian metric $ds^2=g_{\mu\nu}dx^{\mu}dx^{\nu}$, $\mu,\nu=0,1,2,3$, the optical metric can be obtained by impossing the condition  $ds^2=0$. Thus, the optical metric becomes
\begin{equation}
dt^2=g^{OP}_{ij}dx^{i}dx^{j}\,,\,\,\,\,{i,j=1,2,3}\,,
\end{equation}
and which after setting $\theta=\pi/2$ it becomes a two dimensional Riemannian metric
\begin{equation}\label{opticm2d:1}
    dt^2=g^{OP-2d}_{ab}dx^{a}dx^{b}\,,\,\,\,\, a,b=1,2\,.
\end{equation}
We can use the curvatures \eqref{geocurv:1} and \eqref{gausscurv:1} to study geodesic properties of any static spacetime. Indeed, given a static spherically symmetric metric  of the form
\begin{equation}
    ds^2=g_{tt}(r)dt^2+g_{rr}(r)dr^2+g_{\theta\theta}(r)d\theta^2+g_{\phi\phi}(r,\theta)d\phi^2\,,
\end{equation}
then the $2$-dimensional optical metric \eqref{opticm2d:1} can be written
\begin{equation}\label{optic:1}
    dt^2=-\frac{g_{rr}(r)}{g_{tt}(r)}dr^2-\frac{\mathbf{g}_{\phi\phi}(r)}{g_{tt}(r)}d\phi^2\,,
\end{equation}
where $\mathbf{g}_{\phi\phi(r)}=g_{\phi\phi}(r,\theta=\pi/2)$.  A remarkable fact is that the photon orbits, which are of the null type in a spacetime, become spatial geodesics as seen from the optical metric. Thus, the photon sphere of a black hole will become spatial and is entirely characterized by condition \eqref{condgeo:1}.\\
The stability of the photon sphere can be analyzed using the Gaussian curvature of the optical geometry given in  \eqref{condg:2}. See \cite{Qiao:2022jlu,Qiao:2022hfv} for more details.

\subsection{The Jacobi metric and the massive particle surface}
The case of timelike trajectories followed by massive particle surfaces is more subbtle. Instead of the optical metric we have to build an analogous Riemannian metric that inherits the geodesic structure of the spacetime, the Jacobi metric. This metric can be obtained by projecting the spacetime metric over the directions of its Killing vectors \cite{Gibbons:2015qja, Arganaraz:2021fwu}. In static spacetimes the projection is made over the Killing vector $\partial_t$, this projection is equivalent to project the spacetime metric over surfaces of constant energy. The Jacobi metric comming from \eqref{metric:1} is given, in the equatorial plane, by
\begin{equation}\label{jacobim:1}
J_{ij}dx^{i}dx^{j}=\left[-\mathcal{E}^2g^{tt}-m^2\right]\left(g_{rr}dr^2+\mathbf{g}_{\phi\phi}d\phi^2\right)\,.
\end{equation}
When $m=0$ the Jacobi metric \eqref{jacobim:1} reduces, up to a factor $\mathcal{E}^2$, to the optical metric \eqref{opticm2d:1} which is Riemannian and therefore a direct calculation of the Gaussian and geodesic curvatures can be carried out using equations \eqref{gausscurv:1} and \eqref{geocurv:1} respectively. However, now the condition \eqref{condgeo:1} transforms to a new condition for the existence of time-like circular orbits. The criteria for determining its stability remains, the sign of the Gaussian curvature. On the other side, the condition \eqref{geocurv:1} leads to the expression
\begin{equation}\label{cond:1}
\mathbf{g}'_{\phi\phi}F+\mathbf{g}_{\phi\phi}F'=0\,,
\end{equation}
where $F$ is the conformal factor of the Jacobi metric \eqref{jacobim:1}:
\begin{equation}\label{F:1}
F=-\mathcal{E}^2g^{tt}-m^2\,.
\end{equation}
The condition \eqref{cond:1} carries information about the photon spheres ($m=0$) and massive particle surfaces ($m\neq0$).\\
Consider the condition \eqref{cond:1}  with $F$ given by \eqref{F:1} then 
\begin{equation}\label{mpsgen:1}
    \frac{\mathcal{E}^2}{m^2}=\frac{\mathbf{g}'_{\phi\phi}g^2_{tt}}{\mathbf{g}_{\phi\phi}g'_{tt}-\mathbf{g}'_{\phi\phi}g_{tt}}\,.
\end{equation}
When the denominator of \eqref{mpsgen:1} is set to zero $m=0$ we obtain the condition for the existence of a photon sphere. On the other side, if the right side of \eqref{mpsgen:1} is constant then there exists a massive particle surface ($m\neq 0$). Moreover, applying the condition $d\mathcal{E}/dr=0$ the ISCO (Innermost stable circular orbit) calculated.\\
There is a relationship between curvatures when evaluated at circular geodesics. The Gaussian curvature \eqref{gausscurv:1} calculated using the Jacobi metric \eqref{jacobim:1} can be written as a function of the geodesic curvature \eqref{geocurv:1}:
\begin{eqnarray}
    \mathcal{K}&=&-\frac{1}{2\sqrt{UV}}\frac{\partial}{\partial r}\left(\frac{\partial_r V}{\sqrt{UV}}\right)\\
        &=&-\frac{\partial_r \kappa_g}{\sqrt{Fg_{rr}}}\,,
\end{eqnarray}
hence, the sign of $\mathcal{K}$ is determined by the derivative of the geodesic curvature $\kappa_g$, thus \footnote{When $\mathcal{K}=0$ we have marginally stable circular orbit.} 
\begin{eqnarray}
    \text{If}\,\,\partial_r \kappa> 0 \Rightarrow \mathcal{K}< 0,\\
    \text{If}\,\,\partial_r \kappa<0 \Rightarrow \mathcal{K}>0.
\end{eqnarray}
Taking into account equation \eqref{kmps}, we can see that  $\kappa_g(r_o)=\mathcal{K}(r_o)=0$, has to be satisfied by the ISCO, which stands as the boundary between stable and unstable orbits, the so called a marginally stable circular orbit.\\
Setting $g_{tt}=-f(r)$, $g_{rr}(r)=1/g(r)$ and $g_{\phi\phi}=h(r)$, from \eqref{cond:1} we arrive to the results presented in \cite{Bermudez-Cardenas:2024bfi}, namely
\begin{equation}\label{cond:2}
    \frac{\mathcal{E}^2}{m^2}=\frac{h'f^2}{h'f-hf'}.
\end{equation}
An interesting geometrical fact appears when we consider the massive case. Condition \eqref{cond:2} is known  as the master equation \cite{Kobialko:2022uzj,Bogush:2023ojz,Bogush:2024fqj} and if the right side of it is constant then there exist a massive particle surface. Therefore the criteria for the existence of photon spheres and massive particle surfaces is encoded in \eqref{cond:2}. Not only that, the geometry of the corresponding surface is also encoded. The photon sphere is a totally umbilic surface \cite{Claudel:2000yi,Senovilla:2011np,Okumura:1967} while the massive particle surface is partially umbilic, although from the perspective of the Jacobi metric it becomes a totally umbilical surface \cite{Bermudez-Cardenas:2024bfi}. Finally, the ISCO can be also found by $d\mathcal{E}/dr=0$ leading to
\begin{equation}
   3ff'+rff''-2r(f')^{2}=0.
\end{equation}
The formalism described above has been used to study several types of spacetimes providing a potent geometric approach for analyzing the gedesic structure of any spacetime. \cite{Gibbons:2015qja, Das:2016opi,Arganaraz:2019fup,Duenas-Vidal:2022kcx,Bermudez-Cardenas:2025hrp,Bermudez-Cardenas:2025duw,Cunha:2022nyw}.

\appendix




\bibliographystyle{elsarticle-num} 
\bibliography{biblio}

@article{Qiao:2022jlu,
    author = "Qiao, Chen-Kai and Li, Ming",
    title = "{Geometric approach to circular photon orbits and black hole shadows}",
    eprint = "2204.07297",
    archivePrefix = "arXiv",
    primaryClass = "gr-qc",
    doi = "10.1103/PhysRevD.106.L021501",
    journal = "Phys. Rev. D",
    volume = "106",
    number = "2",
    pages = "L021501",
    year = "2022"
}

@article{Qiao:2022hfv,
    author = "Qiao, Chen-Kai",
    title = "{Curvatures, photon spheres, and black hole shadows}",
    eprint = "2208.01771",
    archivePrefix = "arXiv",
    primaryClass = "gr-qc",
    doi = "10.1103/PhysRevD.106.084060",
    journal = "Phys. Rev. D",
    volume = "106",
    number = "8",
    pages = "084060",
    year = "2022"
}

@article{Gibbons:2015qja,
    author = "Gibbons, G. W.",
    title = "{The Jacobi-metric for timelike geodesics in static spacetimes}",
    eprint = "1508.06755",
    archivePrefix = "arXiv",
    primaryClass = "gr-qc",
    doi = "10.1088/0264-9381/33/2/025004",
    journal = "Class. Quant. Grav.",
    volume = "33",
    number = "2",
    pages = "025004",
    year = "2016"
}

@article{Claudel:2000yi,
    author = "Claudel, Clarissa-Marie and Virbhadra, K. S. and Ellis, G. F. R.",
    title = "{The Geometry of photon surfaces}",
    eprint = "gr-qc/0005050",
   archivePrefix = "arXiv",
    doi = "10.1063/1.1308507",
    journal = "J. Math. Phys.",
    volume = "42",
    pages = "818--838",
    year = "2001"
}

@article{Cunha:2022nyw,
    author = "Cunha, Pedro V. P. and Herdeiro, Carlos A. R. and Novo, Jo\~ao P. A.",
    title = "{Null and timelike circular orbits from equivalent 2D metrics}",
    eprint = "2207.14506",
    archivePrefix = "arXiv",
    primaryClass = "gr-qc",
    doi = "10.1088/1361-6382/ac987e",
    journal = "Class. Quant. Grav.",
    volume = "39",
    number = "22",
    pages = "225007",
    year = "2022"
}

@article{Arganaraz:2019fup,
    author = "Arga{\~n}araz, Marcos and Lasso Andino, Oscar",
    title = "{Dynamics in wormhole spacetimes: a Jacobi metric approach}",
    eprint = "1906.11779",
    archivePrefix = "arXiv",
    primaryClass = "gr-qc",
    doi = "10.1088/1361-6382/abcf86",
    journal = "Class. Quant. Grav.",
    volume = "38",
    number = "4",
    pages = "045004",
    year = "2021"
}

@article{Das:2016opi,
    author = "Das, Praloy and Sk, Ripon and Ghosh, Subir",
    title = {{Motion of charged particle in Reissner\textendash{}Nordstr\"om spacetime: a Jacobi-metric approach}},
    eprint = "1609.04577",
    archivePrefix = "arXiv",
    primaryClass = "gr-qc",
    doi = "10.1140/epjc/s10052-017-5295-6",
    journal = "Eur. Phys. J. C",
    volume = "77",
    number = "11",
    pages = "735",
    year = "2017"
}

@article{Duenas-Vidal:2022kcx,
    author = "Duenas-Vidal, {\'A}lvaro and Lasso Andino, Oscar",
    title = "{The Jacobi metric approach for dynamical wormholes}",
    eprint = "2212.14147",
    archivePrefix = "arXiv",
    primaryClass = "gr-qc",
    doi = "10.1007/s10714-022-03060-w",
    journal = "Gen. Rel. Grav.",
    volume = "55",
    number = "1",
    pages = "9",
    year = "2023"
}

@article{Bermudez-Cardenas:2024bfi,
    author = "Berm{\'u}dez-C{\'a}rdenas, Boris and Andino, Oscar Lasso",
    title = "{Massive particle surfaces, partial umbilicity, and circular orbits}",
    eprint = "2409.10789",
    archivePrefix = "arXiv",
    primaryClass = "gr-qc",
    doi = "10.1103/PhysRevD.111.064001",
    journal = "Phys. Rev. D",
    volume = "111",
    number = "6",
    pages = "064001",
    year = "2025"
}

@article{Bermudez-Cardenas:2025duw,
    author = "Berm{\'u}dez-C{\'a}rdenas, Boris and Andino, Oscar Lasso",
    title = "{Massive particle surfaces and black hole shadows from intrinsic curvature}",
    eprint = "2503.21203",
    archivePrefix = "arXiv",
    primaryClass = "gr-qc",
    doi = "10.1140/epjc/s10052-025-15009-9",
    journal = "Eur. Phys. J. C",
    volume = "85",
    number = "11",
    pages = "1266",
    year = "2025"
}

@article{Bermudez-Cardenas:2025hrp,
    author = "Berm{\'u}dez-C{\'a}rdenas, Boris and Lasso Andino, Oscar",
    title = "{Light rings, timelike circular orbits and curvature of traversable wormholes}",
    journal = "ArXiv",
    eprint = "2504.10732",
    archivePrefix = "arXiv",
    primaryClass = "gr-qc",
    month = "4",
    year = "2025"
}

@article{LassoAndino:2026new,
    author = "Lasso Andino, Oscar and Le{\'o}n-Arteaga, Axel and Ram{\'\i}rez-Ulloa, Guillermo",
    title = "{Photon spheres and bulk probes in $\text{AdS}_3$/$\text{CFT}_2$: the quantum BTZ black hole}",
    journal = "ArXiv",
    eprint = "2603.09169",
    archivePrefix = "arXiv",
    primaryClass = "hep-th",
    month = "3",
    year = "2026"
}

@article{Senovilla:2011np,
 author = "J. M. M. Senovilla",
title = "{Umbilical-Type Surfaces in Spacetime}",
eprint = "1111.6910",
year = "2011",
journal = "ArXiv",
primaryClass = "math.DG"
}

@article{Okumura:1967,
author = "M. Okumura",
title = "{Totally umbilical hypersurfaces of a locally product Riemannian manifold}",
journal = "Kodai Math. Sem. Rep.",
vol = "19",
issue = "1",
Pages = "35 - 42",
year = "1967",
doi = "10.2996/kmj/1138845339"
}

@article{Bogush:2023ojz,
    author = "Bogush, Igor and Kobialko, Kirill and Gal'tsov, Dmitri",
    title = "{Massive particle surfaces}",
    doi = "10.1103/PhysRevD.108.044070",
    journal = "Phys. Rev. D",
    volume = "108",
    number = "4",
    pages = "044070",
    year = "2023"
}

@article{Kobialko:2022uzj,
    author = "Kobialko, Kirill and Bogush, Igor and Gal'tsov, Dmitri",
    title = "{Geometry of massive particle surfaces}",
    eprint = "2208.02690",
    archivePrefix = "arXiv",
    primaryClass = "gr-qc",
    doi = "10.1103/PhysRevD.106.084032",
    journal = "Phys. Rev. D",
    volume = "106",
    number = "8",
    pages = "084032",
    year = "2022"
}

@article{Bogush:2024fqj,
    author = "Bogush, Igor and Kobialko, Kirill and Gal'tsov, Dmitri",
    title = "{Constructing massive particles surfaces in static spacetimes}",
    eprint = "2402.03266",
    archivePrefix = "arXiv",
    primaryClass = "gr-qc",
    doi = "10.1140/epjc/s10052-024-12751-4",
    journal = "Eur. Phys. J. C",
    volume = "84",
    number = "4",
    pages = "387",
    year = "2024"
}

@article{Arganaraz:2021fwu,
    author = "Arga\~naraz, Marcos A. and Andino, Oscar Lasso",
    title = "{A Riemannian geometric approach for timelike and null spacetime geodesics}",
    eprint = "2112.10910",
    archivePrefix = "arXiv",
    primaryClass = "gr-qc",
    doi = "10.1007/s10714-024-03314-9",
    journal = "Gen. Rel. Grav.",
    volume = "56",
    pages = "121",
    year = "2024"
}

@book{Berg:1988,
author = "M. Berger, B. Gostiaux", 
title = "{Differential Geometry: Manifolds, Curves and Surfaces.}",
editorial= "New York: Springer",
publisher= "New York: Springer",
pages = "p. 416", 
year = "1988."
}

@book{Kre:1991,
author = "E. Kreyszig", 
title = "Differential Geometry",
editorial="New York: Dover",
publisher="New York: Dover",
pages = "131", 
year = "1991"
}

@article{Qiao:2025ojr,
    author = "Qiao, Chen-Kai and Su, Ping and Huang, Yang",
    title = "{A general discussion on photon spheres in different categories of spacetimes}",
    eprint = "2602.04573",
    archivePrefix = "arXiv",
    primaryClass = "gr-qc",
    reportNumber = "ChinaXiv:202412.00090",
    doi = "10.1140/epjc/s10052-025-14435-z",
    journal = "Eur. Phys. J. C",
    volume = "85",
    number = "6",
    pages = "709",
    year = "2025"
}

@article{Vertogradov:2024dpa,
    author = {Vertogradov, Vitalii and {\"O}vg{\"u}n, Ali},
    title = "{General approach on shadow radius and photon spheres in asymptotically flat spacetimes and the impact of mass-dependent variations}",
    eprint = "2404.18536",
    archivePrefix = "arXiv",
    primaryClass = "gr-qc",
    doi = "10.1016/j.physletb.2024.138758",
    journal = "Phys. Lett. B",
    volume = "854",
    pages = "138758",
    year = "2024"
}

@article{Bardeen1968,
  author    = "Bardeen, J. M.",
  title     = "{Non-singular general-relativistic gravitational collapse}",
   journal = {Proceedings of the International Conference GR5},
  year      = "1968",
  address   = "Tbilisi, USSR",
  pages     = "174"
}

@article{EventHorizonTelescope:2019dse,
    author = "Akiyama, Kazunori and others",
    collaboration = "Event Horizon Telescope",
    title = "{First M87 Event Horizon Telescope Results. I. The Shadow of the Supermassive Black Hole}",
    eprint = "1906.11238",
    archivePrefix = "arXiv",
    primaryClass = "astro-ph.GA",
    doi = "10.3847/2041-8213/ab0ec7",
    journal = "Astrophys. J. Lett.",
    volume = "875",
    pages = "L1",
    year = "2019"
}

@article{EventHorizonTelescope:2022wkp,
    author = "Akiyama, Kazunori and others",
    collaboration = "Event Horizon Telescope",
    title = "{First Sagittarius A* Event Horizon Telescope Results. I. The Shadow of the Supermassive Black Hole in the Center of the Milky Way}",
    eprint = "2311.08680",
    archivePrefix = "arXiv",
    primaryClass = "astro-ph.HE",
    doi = "10.3847/2041-8213/ac6674",
    journal = "Astrophys. J. Lett.",
    volume = "930",
    number = "2",
    pages = "L12",
    year = "2022"
}

@article{Luminet:1979nyg,
    author = "Luminet, J. -P.",
    title = "{Image of a spherical black hole with thin accretion disk}",
    journal = "Astron. Astrophys.",
    volume = "75",
    pages = "228--235",
    year = "1979"
}

@article{Falcke:1999pj,
    author = "Falcke, Heino and Melia, Fulvio and Agol, Eric",
    title = "{Viewing the shadow of the black hole at the galactic center}",
    eprint = "astro-ph/9912263",
    archivePrefix = "arXiv",
    reportNumber = "HFA-EPRINT-NO-33",
    doi = "10.1086/312423",
    journal = "Astrophys. J. Lett.",
    volume = "528",
    pages = "L13",
    year = "2000"
}

@article{Vagnozzi:2022moj,
    author = "Vagnozzi, Sunny and others",
    title = "{Horizon-scale tests of gravity theories and fundamental physics from the Event Horizon Telescope image of Sagittarius A}",
    eprint = "2205.07787",
    archivePrefix = "arXiv",
    primaryClass = "gr-qc",
    reportNumber = "UCI-HEP-TR-2022-07",
    doi = "10.1088/1361-6382/acd97b",
    journal = "Class. Quant. Grav.",
    volume = "40",
    number = "16",
    pages = "165007",
    year = "2023"
}

@article{Perlick:2021aok,
    author = "Perlick, Volker and Tsupko, Oleg Yu.",
    title = "{Calculating black hole shadows: Review of analytical studies}",
    eprint = "2105.07101",
    archivePrefix = "arXiv",
    primaryClass = "gr-qc",
    doi = "10.1016/j.physrep.2021.10.004",
    journal = "Phys. Rept.",
    volume = "947",
    pages = "1--39",
    year = "2022"
}

@article{Ayzenberg:2023hfw,
    author = "Ayzenberg, D. and others",
    title = "{Fundamental physics opportunities with future ground-based mm/sub-mm VLBI arrays}",
    eprint = "2312.02130",
    archivePrefix = "arXiv",
    primaryClass = "astro-ph.HE",
    doi = "10.1007/s41114-025-00057-0",
    journal = "Living Rev. Rel.",
    volume = "28",
    number = "1",
    pages = "4",
    year = "2025",
    note = "[Erratum: Living Rev.Rel. 28, 7 (2025)]"
}

@article{Lupsasca:2024xhq,
    author = "Lupsasca, Alexandru and C{\'a}rdenas-Avenda{\~n}o, Alejandro and Palumbo, Daniel C. M. and Johnson, Michael D. and Gralla, Samuel E. and Marrone, Daniel P. and Galison, Peter and Tiede, Paul and Keeble, Lennox",
    title = "{The Black Hole Explorer: photon ring science, detection, and shape measurement}",
    eprint = "2406.09498",
    archivePrefix = "arXiv",
    primaryClass = "gr-qc",
    doi = "10.1117/12.3019437",
    journal = "Proc. SPIE Int. Soc. Opt. Eng.",
    volume = "13092",
    pages = "130926Q",
    year = "2024"
}

@article{Gralla:2019xty,
    author = "Gralla, Samuel E. and Holz, Daniel E. and Wald, Robert M.",
    title = "{Black Hole Shadows, Photon Rings, and Lensing Rings}",
    eprint = "1906.00873",
    archivePrefix = "arXiv",
    primaryClass = "astro-ph.HE",
    doi = "10.1103/PhysRevD.100.024018",
    journal = "Phys. Rev. D",
    volume = "100",
    number = "2",
    pages = "024018",
    year = "2019"
}

@article{Wong:2024gph,
    author = "Wong, George N. and Medeiros, Lia and C{\'a}rdenas-Avenda{\~n}o, Alejandro and Stone, James M.",
    title = "{Measuring Black Hole Light Echoes with Very Long Baseline Interferometry}",
    eprint = "2410.10950",
    archivePrefix = "arXiv",
    primaryClass = "astro-ph.HE",
    doi = "10.3847/2041-8213/ad8650",
    journal = "Astrophys. J. Lett.",
    volume = "975",
    number = "2",
    pages = "L40",
    year = "2024"
}

@article{Gralla:2020srx,
    author = "Gralla, Samuel E. and Lupsasca, Alexandru and Marrone, Daniel P.",
    title = "{The shape of the black hole photon ring: A precise test of strong-field general relativity}",
    eprint = "2008.03879",
    archivePrefix = "arXiv",
    primaryClass = "gr-qc",
    doi = "10.1103/PhysRevD.102.124004",
    journal = "Phys. Rev. D",
    volume = "102",
    number = "12",
    pages = "124004",
    year = "2020"
}

@article{Olejak:2019pln,
    author = "Olejak, A. and Belczynski, K. and Bulik, T. and Sobolewska, M.",
    title = "{Synthetic catalog of black holes in the Milky Way}",
    eprint = "1908.08775",
    archivePrefix = "arXiv",
    primaryClass = "astro-ph.SR",
    doi = "10.1051/0004-6361/201936557",
    journal = "Astron. Astrophys.",
    volume = "638",
    pages = "A94",
    year = "2020"
}

@article{Mroz:2024mse,
    author = "Mr{\'o}z, Przemek and others",
    title = "{No massive black holes in the Milky Way halo}",
    eprint = "2403.02386",
    archivePrefix = "arXiv",
    primaryClass = "astro-ph.GA",
    doi = "10.1038/s41586-024-07704-6",
    journal = "Nature",
    volume = "632",
    number = "8026",
    pages = "749--751",
    year = "2024"
}

@article{LIGOScientific:2020kqk,
    author = "Abbott, R. and others",
    collaboration = "LIGO Scientific, Virgo",
    title = "{Population Properties of Compact Objects from the Second LIGO-Virgo Gravitational-Wave Transient Catalog}",
    eprint = "2010.14533",
    archivePrefix = "arXiv",
    primaryClass = "astro-ph.HE",
    reportNumber = "LIGO-P2000077",
    doi = "10.3847/2041-8213/abe949",
    journal = "Astrophys. J. Lett.",
    volume = "913",
    number = "1",
    pages = "L7",
    year = "2021"
}

@article{LIGOScientific:2016dsl,
    author = "Abbott, B. P. and others",
    collaboration = "LIGO Scientific, Virgo",
    title = "{Binary Black Hole Mergers in the first Advanced LIGO Observing Run}",
    eprint = "1606.04856",
    archivePrefix = "arXiv",
    primaryClass = "gr-qc",
    reportNumber = "LIGO-P1600088",
    doi = "10.1103/PhysRevX.6.041015",
    journal = "Phys. Rev. X",
    volume = "6",
    number = "4",
    pages = "041015",
    year = "2016",
    note = "[Erratum: Phys.Rev.X 8, 039903 (2018)]"
}






\end{document}